\documentclass[article,twocolumn,superscriptaddress,showpacs,amsfonts,amsmath,amssymb,showkeys,floatfix]{revtex4}

 \usepackage{graphicx}

\usepackage{amssymb}

\begin{document}
\title{Continuous spin reorientation in antiferromagnetic films}
\author{Juan J. Alonso}
\affiliation{F\'{\i}sica Aplicada I, Universidad de M\'alaga,
29071-M\'alaga, Spain}
\email[E-mail address: ] {jjalonso@Darnitsa.Cie.Uma.Es}
\author{Julio F. Fern\'andez}
\affiliation{ICMA, CSIC and Universidad de Zaragoza, 50009-Zaragoza, Spain}
\email[E-mail address: ] {JFF@Pipe.Unizar.Es}

\begin{abstract}
We study anisotropic antiferromagnetic one-layer films with dipolar and 
nearest-neighbor exchange interactions. We obtain a unified 
phase diagram as a function of effective uniaxial $D_e$ 
and quadrupolar $C$ anisotropy constants. We study in some detail 
how spins reorient continuously below a temperature $T_s$ as 
$T$ and $D_e$ vary.
\end{abstract}

\maketitle


Considerable attention has been devoted to the magnetic properties
of ultrathin magnetic films in the last years. 
An interesting  feature of magnetic films is
discontinous spin reorientation (DSR), i.e., thermally driven 
switching between perpendicular and in-plane spin alignment at a 
temperature $T_r$ below the ordering temperature $T_0$. 
DSR has been studied both experimentally and theoretically. 
It has been established that DSR depends on
the competition between dipolar interactions 
and uniaxial anisotropy often found in films. Continuous spin reorientation
(CSR) is also very interesting. 
A thermally driven CSR transition was first observed
experimentally in bulk systems at some temperature $T_s$ well 
below $T_0$\cite{gyor}.
Below $T_s$, all spins rotate continuously as a whole as
$T$ is varied in these systems.  After these experiments
Horner and Varma\cite{var} proposed an early 
phenomenological model in which
higher anisotropes, which compete with the
uniaxial anisotropy, were required for obtaining  
CSR.  More recently, CSR has been observed in ferromagnetic thin
films \cite{san}. Some nonhomogeneus multilayer
models have been proposed to explain CSR\cite{usadel}.

\begin{figure}[!b]
\includegraphics*[width=80mm]{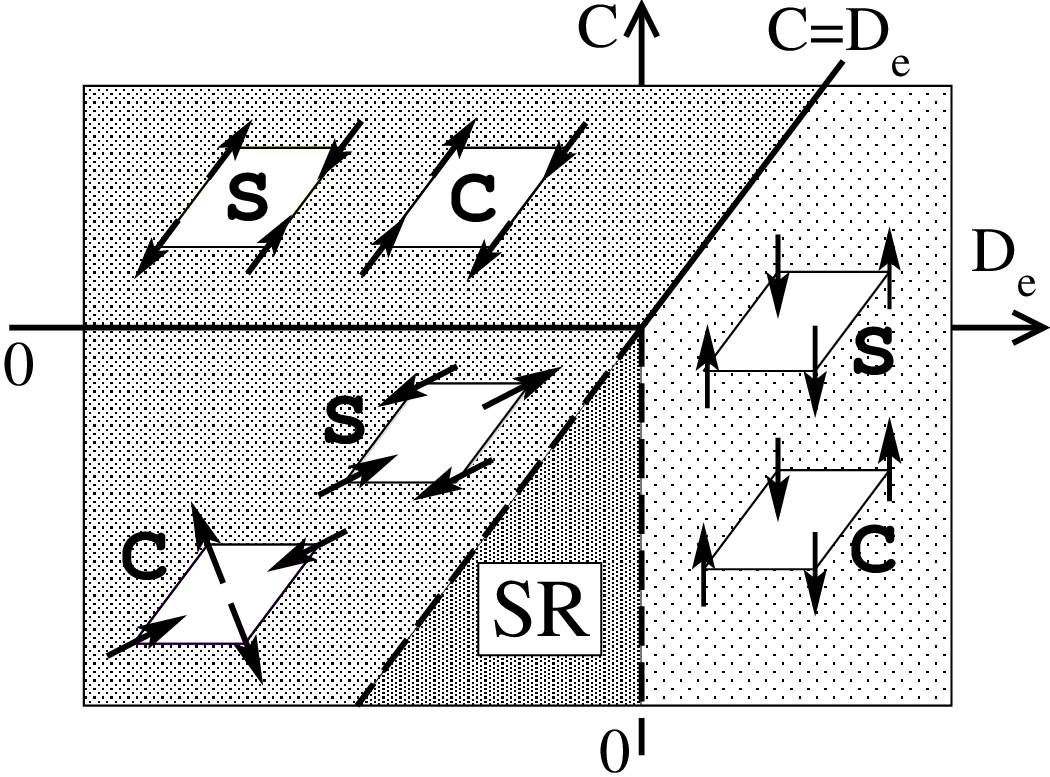}
\caption{Magnetic phase diagram for dipolar
systems in their ground states for $J\leq 0$.
$\mathfrak s$ and $\mathfrak c$ states correspond to
exchange dominated systems with $J<-1.61\varepsilon_d$
and dipolar dominated systems with $-1.61\varepsilon_d<J<0$,
respectively. Full and dashed thick lines stand for first--
and second--order transitions, respectively. SR stands for the
spin reorientation phase, in which $0<\theta<\pi/2$. }
\label{fig1}
\end{figure}

The aim of this paper is to report numerical results
on CSR for one-layer antiferromagnetic films.
It is important to note that CSR is always associated with 
a SR phase defined by its own broken symmetries\cite{landau,ours}. 
For instance, the order parameter $\bf m$ may be perpendicular ($\theta=0$)
to the film plane in the $T_s<T<T_0$ range, 
and tilt away ($0<\theta<\pi$) from the easy magnetization axis below 
$T_s$, thus breaking additional symmetries\cite{ours}.
To our knowledge, such SR magnetic phase has not  
been observed in numerical simulations of antiferromagnetic
films.

\begin{figure}[!t]
\includegraphics*[width=80mm]{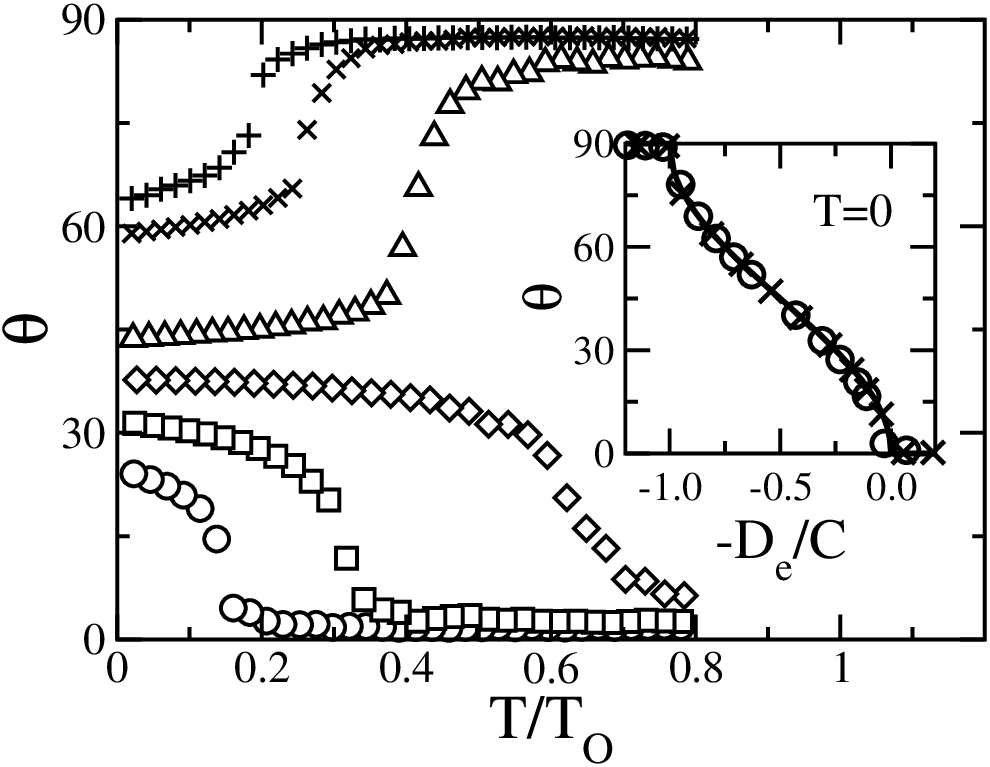}
\caption{Tilt angle $\theta$ versus $T/T_0$ for a pure dipolar system $(J=0)$ of $32 \times 32 \times 1$ spins with $C=-\varepsilon_d$. The system
is ordered for $T<T_0$. Curves correspond from top to bottom to $D_e/C=0.9, 0.8, 0.7, 0.6, 0.5$, and $0.4$ respectively. Similar plots have been obtained
for $J<0$. In the inset, $\theta$ versus $-D_e/C$ for $T=0$ is shown. Solid line corresponds to $tan\theta=\sqrt{D_e/(C-D_e)}$.
$\times$ ($\circ$) stand for systems of $32 \times 32$ spins
with  $J=0$  and variable $D$ ($D=0$ and variable $J$), respectively.
.}
\label{fig2}
\end{figure}

We consider a system of classical unit spins ${\{\bf S}_i\}$ in a
square lattice with hamiltonian $H=H_J+H_d+H_a$ where $H_J$, $H_d$, 
and $H_a$ are for short range exchange, long range dipolar,
and anisotropy interactions, respectively. We use periodic
boundary conditions. The exchange and dipolar energy between 
two antiparallel out of plane nearest neighbors 
spins is $J<0$ and $-\varepsilon_d$, respectively.
Furthermore, there is site uniaxial $-D(S_i^z)^2$
plus fourfold $-C[(S_i^x)^4+(S_i^y)^4]$ anisotropy
energies. Similar
models have been studied for $C=0$, and 
some DSR between degenerate in-plane and out-of-plane states
have been found as $T, J$ and $D$ vary\cite{isaac}.
For ferromagnetic films, on the other hand, Jensen et al. have 
considered $C\ne0$ but obtained \cite{benne} phase diagrams only for 
$T=0$.

Ground state configurations obtained from Monte Carlo simulations 
are exhibited in Fig.~1. States designated with a $\mathfrak c$, 
can be defined \cite{ours} by 
$S_i^z=\tau_i^{z}\cos \theta,\; S_i^y=\tau_i^{y}\sin \theta \sin\phi,\;
S_i^x=\tau_i^{x}\sin \theta \cos \phi$
where $\tau_i^x=(-1)^{y(i)}$, $\tau_i^y=(-1)^{x(i)}$,
and $\tau_i^z=(-1)^{x(i)+y(i)}$. On the other hand,
states designated with a $\mathfrak s$ are defined by
$\tau_i^{x}=\tau_i^y=\tau_i^z=(-1)^{x(i)+y(i)}$. A suitable
order parameter for both $\mathfrak c$ and $\mathfrak s$ is $
m^\alpha =N^{-1}\sum_i  S_i^\alpha \tau^{\alpha}_i$.

We calculate energies for these configurations as in Ref.~[6], and
find a surface anisotropy energy $\Delta$
that behaves as an easy axis anisotropy. 
We find that $\Delta=2\varepsilon_d$ for $\mathfrak s$ states and 
$\Delta=-1.23\varepsilon_d+2J$ for
$\mathfrak c$ states.
This suggests we can define an effective anisotropy 
as $D_e=D+\Delta$ and obtain a unified phase diagram for
both  $\mathfrak s$ and $\mathfrak c$ states 
for $T=0$, as shown in Fig.~1. $\mathfrak s$ configurations give 
a lower energy for $J<-1.61\varepsilon_d$ while 
$\mathfrak c$ states are more favorable for $-1.61\varepsilon_d<J<0$.
Interesting experimental realizations of the latter condition could be
found in Ref.~[9].

In both cases we find a z-collinear phase ($\theta=0$) if both $D_e>0$ and 
$D_e>C$ are fulfilled. In-plane configurations ($\theta=\pi/2$) are prefered
for $D_e<C$. More interestingly, we obtain a spin reorientation phase 
for $C<D_e<0$ in which $\theta$ covers the $0<\theta<\pi/2$ range.
Minimization of the total energy gives $tan\theta=\sqrt{D_e/(C-D_e)}$
and therefore spins rotate continuously from $\theta=0$ to $\pi/2$
as $D_e$ varies from $0$ to $C$, as shown in the inset of Fig.~2. 
Symbols in the same inset correspond to numerical data obtained by cooling
from the paramagnetic phase to $T<<\varepsilon_d, J$ for different
values of $D_e/C$. 

\begin{figure}[!t]
\includegraphics*[width=80mm]{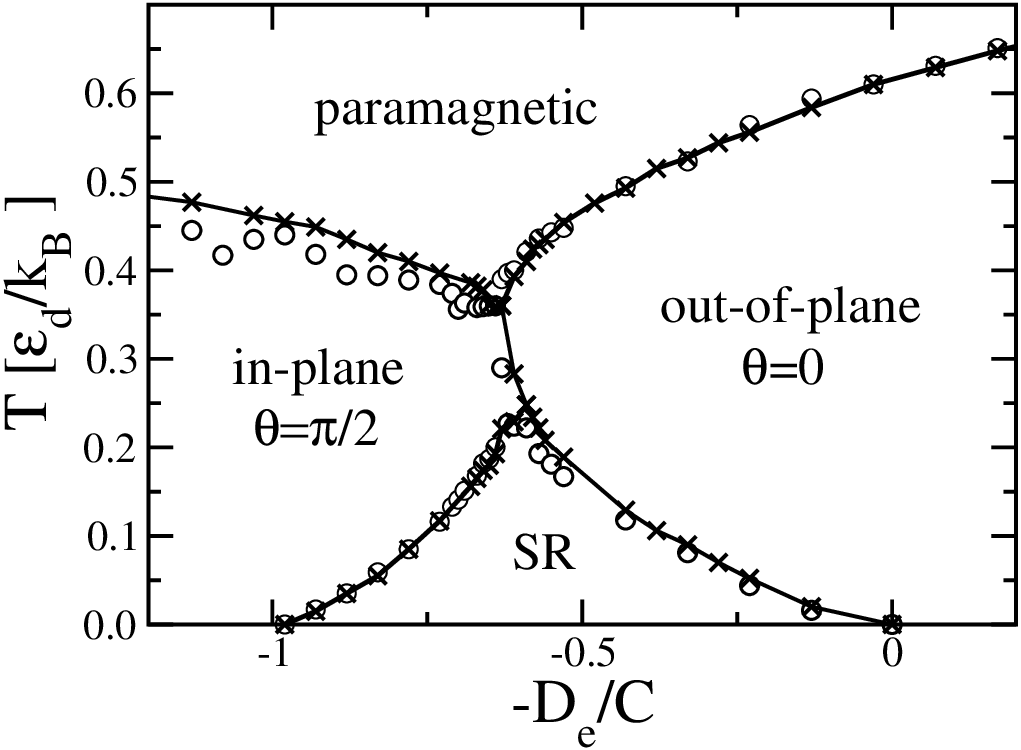}
\caption{Phases of pure dipolar ($J=0$)
films for $C=-\varepsilon_d$ obtained from MC simulations.
$\circ$ ($\times$) stand for systems of $16\times 16$ ($32\times 32$) spins
respectively. Transition temperatures have been
obtained from peaks observed in the specific heat. Systems
have been cooled in $\Delta T =-0.01 \varepsilon_d/k_B$ steps
of $4 \times 10^5$ MC sweeps each. Similar phase diagrams have
been obtained for $J<0$.}
\label{fig3}
\end{figure}

We have explored temperature driven CSR in the SR phase by MC simulations for
both exchange and dipolar dominated systems. 
For that purpose we calculate the order parameter $(m^x, m^y, m^z)$ 
and the energy as a function of $T$. 
We find two different regions (see figs.~2 and 3). Upon cooling 
below $T_0$, in-plane configurations appear for $D_e/C>0.65$, and
spins rotate towards the $z$ axis below a second order 
transition at $T_s<T_0$. On the other hand, for $D_e/C<0.65$ spins 
point out of the plane below $T_0$, and rotate 
towards the $xy$ plane as $T$ decreases below $T_s$. Finally, we find, 
that the ratio $T_s/T_0$, as in $d=3$ systems\cite{ours}, seems to 
depend mainly on $D_e/C$ and not on $J$ or $\varepsilon_d$. 


\begin{references}
\bibitem{gyor} E. M. Gyorgy, J. P. Remeika and F. B. Hagedorn,
J. Appl. Phys. {\bf 39}, 1369 (1968).
\bibitem{var} H. Horner and C. M. Varma, Phys. Rev.
Lett. {\bf 20}, 845 (1968).
\bibitem{san} M. Farle et al., Phys Rev. B {\bf 55}, 3708 (1997);
G. Garreau, E. Beaurepaire, K. Oujnadela and M. Farle, Phys. Rev. B
{\bf 53}, 1083 (1996); R. Sellmann et al., Phys. Rev. Lett. {\bf 64},
054418 (2001).
\bibitem{usadel} A. Moshel and K. D. Usadel,  Phys. Rev. B {\bf 51},
16111 (1995); L. Udvardi et al., Philos. Mag. B  {\bf 81}, 613 (2001).
\bibitem{landau} L. D. Landau and E. M. Lifshitz, {\it 
Electrodynamics of Continuous Media}, 2nd ed. (Pergamon, Oxford, 2004),
pp. 159-162.
\bibitem{ours}
J. F. Fern\'andez and J. J. Alonso, Phys. Rev. B {\bf 73}, 024412 (2006)
\bibitem{isaac} 
K. De'Bell, A. B. MacIsaac, J. P. Whitehead, Rev. Mod. Phys.
{\bf 72}, 225 (2000) and references therein.
\bibitem{benne}
P. J. Jensen and K. H. Bennemann
Phys. Rev. {\bf 42}, 849 (1990).
\bibitem{dipo}G. Ahlers, A. Kornblit, and H. J. Guggenheim,
Phys. Rev. Lett. {\bf 34}, 1227 (1975); G. Mennenga, L. J. de Jongh,
and W. J. Huiskamp, J. Magn. Magn. Mater. {\bf 44}, 59
(1984); M. R. Roser and L. R. Corruccini, Phys. Rev. Lett.{\bf 65},
1064 (1990);
D. Bitko, T. F. Rosenbaum, and G. Aeppli, Phys.
Rev. Lett. {\bf 77}, 940 (1996).

\end{references}
\end{document}